# IRCI Free MIMO-OFDM SAR Using Circularly Shifted Zadoff-Chu Sequences

Yun-He Cao[1], *Member*, *IEEE*, and Xiang-Gen Xia[1,2], *Fellow*, *IEEE*

**Abstract**

Cyclic prefix (CP) based MIMO-OFDM radar has been recently proposed for distributed transmit antennas, where there is no inter-range-cell interference (IRCI). It can collect full spatial diversity and each transmitter transmits signals with the same frequency band, i.e., the range resolution is not reduced. However, it needs to transmit multiple OFDM pulses consecutively to obtain range profiles for a single swath, which may be too long in time for a reasonable swath width. In this letter, we propose a CP based MIMO-OFDM synthetic aperture radar (SAR) system, where each transmitter transmits only a single OFDM pulse to obtain range profiles for a swath and has the same frequency band, thus the range resolution is not reduced. It is IRCI free and can collect the full spatial diversity if the transmit antennas are distributed. Our main idea is to use circularly shifted Zadoff-Chu sequences as the weighting coefficients in the OFDM pulses for different transmit antennas and apply spatial filters with multiple receive antennas to divide the whole swath into multiple subswaths, and then each subswath is reconstructed/imaged using our proposed IRCI free range reconstruction method.

**Index Terms**

MIMO-OFDM, synthetic aperture radar (SAR), cyclic prefix (CP), circularly shifted Zadoff-Chu sequences, inter-range-cell interference (IRCI) free.

This work was supported in part by the National Natural Science Foundation of China (61372136, 61172137), the Fundamental Research Funds for the Central Universities (K5051202005, K5051302089), the Air Force Office of Scientific Research (AFOSR) under Grant FA9550-12-1-0055, and by the China Scholarship Council (CSC).

[1] National Laboratory of Radar Signal Processing, Xidian University, Xi'an, P.R. China, 710071. (E-mail: cyh_xidian@163.com; xxia@ee.udel.edu).

[2] Department of Electrical and Computer Engineering, University of Delaware, Newark, DE 19716, USA. (Email: xxia@ee.udel.edu).

## I. INTRODUCTION

Multiple input/transmit and multiple output/receive (MIMO) radar has attracted much attention in recent years, see, for example, [1]-[3]. It is recognized that distributed MIMO radar may be able to collect the spatial diversity, while co-located MIMO radar is able to improve the detection capability of moving targets. For distributed MIMO radar to collect the spatial diversity, the signal waveforms transmitted from the multiple transmitters have to be orthogonal despite their delays in the time domain, which is usually challenging to design, particularly in synthetic aperture radar (SAR) applications with high resolution in both range and azimuth. With the current MIMO SAR waveform designs, to collect the full spatial diversity, the multiple waveforms have non-overlapping frequency bands, which reduces the range resolution. On the other hand, if the range resolution is not reduced, all the multiple waveforms need to have the same frequency band but in this case, their delayed versions in the time domain may not be orthogonal each other and thus the full spatial diversity may not be achieved in distributed MIMO SAR. Despite the challenge, a short-term shift-orthogonal chirp waveform combined with digital beamforming (DBF) on receive in elevation is proposed [4] and four chirp modulation diversity waveforms are proposed in [5].

Orthogonal frequency division multiplexing (OFDM) waveforms have been recently studied for SAR applications [6]-[8]. In order to obtain a group of orthogonal OFDM waveforms suitable for MIMO SAR imaging, an interleaving frame structure in the frequency domain is proposed [6], which divides the available bandwidth into several non-overlapping subbands. All the aforementioned SAR imaging schemes use the traditional matched filter, which will result in range sidelobes. Using a cyclic prefix (CP) at the transmitter and a specially proposed range reconstruction (not necessary matched filter) at the receiver, the OFDM SAR imaging proposed in [8] for single channel transmitter achieves ideally zero sidelobes in range called inter-range-cell interference (IRCI) free. A CP based MIMO-OFDM radar for distributed antennas is proposed in [9], where all the waveforms have the same frequency band, i.e., the range resolution is not reduced, with IRCI free range reconstruction, while the full spatial diversity is achieved. In addition to the CP insertion at the transmitter and the newly proposed IRCI-free range reconstruction (not matched filtering) at the receiver, the key idea for the MIMO-OFDM waveform design in [9] is that the designed multiple OFDM waveforms are orthogonal in the discrete frequency domain, which is not affected by delays in the time domain. However, in order to achieve the IRCI-free range reconstruction, multiple (at least the number of transmit antennas) consecutive OFDM pulses need to be transmitted for a single swath, which may be too long in time for a reasonable swath width.

In this letter, we propose a CP based MIMO-OFDM SAR system, where each transmitter transmits only a single OFDM pulse to obtain range profiles for a single swath and has the same frequency band, thus the range resolution is not reduced. It is IRCI free, i.e., there is no IRCI from any transmitter (or channel), and can collect the full spatial diversity if the transmit antennas are distributed. Our main idea is to use circularly shifted Zadoff-Chu sequences [11], [12] as the weighting coefficients in the OFDM pulses for different transmit antennas and apply spatial filters with multiple receive antennas to divide the whole swath into multiple subswaths, and then each subswath is reconstructed/imaged using our proposed IRCI free range reconstruction method. Note that a CP based MIMO-OFDM radar for (only) co-located transmitters is proposed in [10] that also achieves IRCI free range reconstruction. This letter is mainly for distributed transmitters.

The remaining of this letter is organized as follows. Section II introduces the transmit and receive signal model for MIMO SAR with CP based OFDM waveforms. Section III proposes our IRCI free range reconstruction method for MIMO SAR system and the requirements for the multiple transmitted OFDM waveforms. Section IV designs circularly shifted Zadoff-Chu sequences to satisfy the requirements of the proposed scheme for MIMO SAR. Section V presents some simulation results. At last, Section VI concludes this letter.

## II. MIMO-OFDM SAR SIGNAL MODEL

Consider a MIMO-OFDM SAR imaging system with multiple transmit and multiple receive antennas. We use CP based OFDM sequences as transmit waveforms of the MIMO-OFDM SAR. The diagram of MIMO SAR transmit OFDM waveforms with CP is shown in Fig. 1 where we assume that the multiple transmit antennas are distributed although the method below also applies to the case of co-located transmit antennas. Multiple complex-valued weighting sequences for subcarriers with length $N$ are adopted. The waveform bandwidth is $B = N\Delta f$, where $\Delta f$ represents the frequency difference between two adjacent subcarriers.

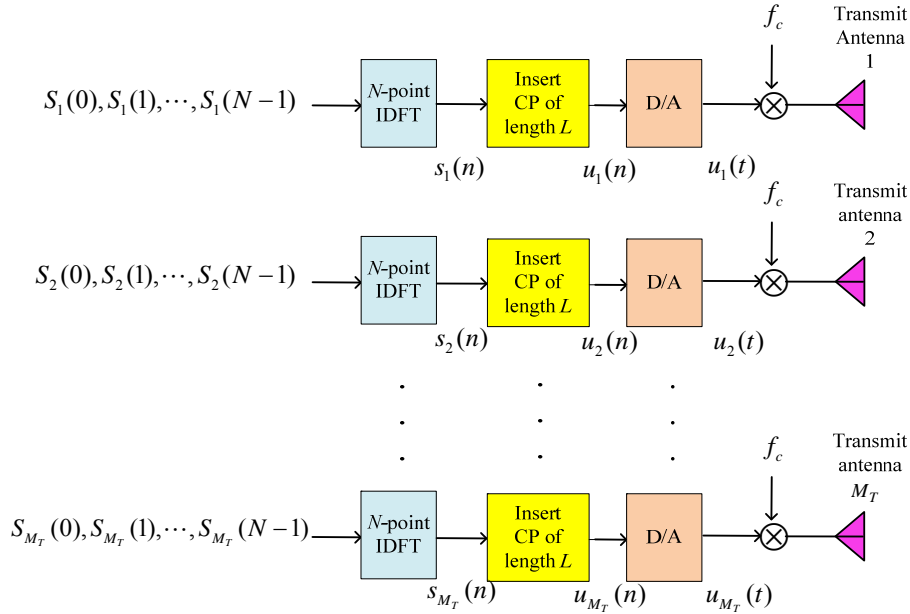

Fig. 1. MIMO-OFDM SAR transmitters.

The discrete time domain OFDM waveform can be obtained by the $N$-point inverse discrete Fourier transform (IDFT) as follows

$$s_m(n) = \frac{1}{\sqrt{N}} \sum_{k=0}^{N-1} S_m(k) \exp\left(\frac{j2\pi nk}{N}\right). \tag{1}$$

We add a CP at the beginning of the transmit waveform and the CP length should be larger than the maximum range cell number in a swath (or an effective swath in this paper) [8] to suppress the range sidelobes. In order to minimize the CP length so as to reduce the unnecessary transmission energy and also for convenience, we let the CP length equal to the maximum range cell number in an effective swath that will be specified later. By adding the CP at the beginning of the waveform, one guarantees that the received signal has a full period of the transmitted waveform symbol for each range cell after removing the CP part at the receiver. Thus, in Fig. 1,

$$u_m(n) = \begin{cases} s_m(n+N-L), & 0 \leq n < L \\ s_m(n-L), & L \leq n < L+N-1 \end{cases}. \tag{2}$$

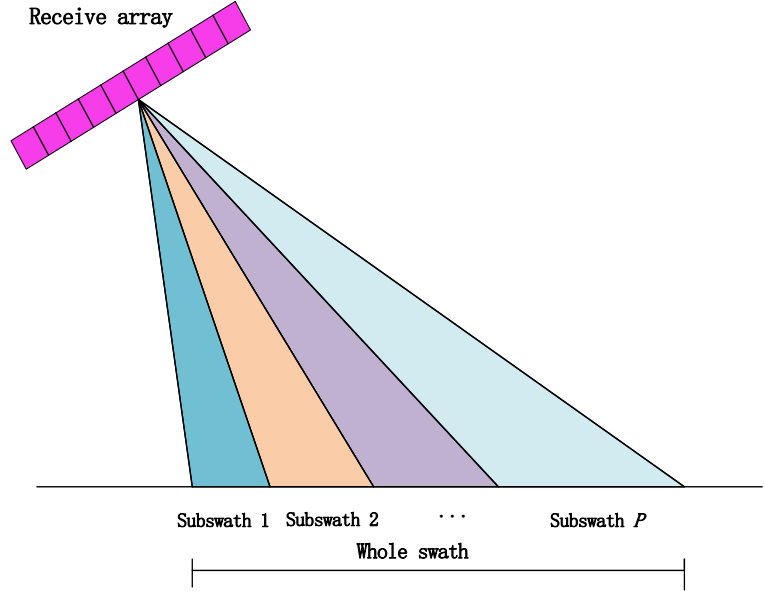

Fig. 2. Receive MIMO SAR with spatial filters.

Assume that the receive array is co-located and multiple receive digital beams (spatial filters) are formed simultaneously [4], [6]. The whole swath is divided into $P$ subswaths by narrow receive beams so that the echo length of each subswath is significantly reduced. The receive MIMO SAR with spatial filter diagram is shown in Fig. 2. Assume that the occupied range cell number of the $p$th subswath is $L_p$ and the maximum range cell number $L_o$ of all the subswaths satisfies

$$L_o \leq N/M_T, \tag{3}$$

where $L_o = \max(L_1, L_2, \cdots, L_P)$. The CP length $L$ for each transmit OFDM waveform is $L = L_o$.

After analog to digital (A/D) sampling with sampling frequency $f_s = B$ and the spatial filtering, the discrete baseband received signal of the $p$th subswath can be written as

$$r_p(n) = \sum_{m=1}^{M_T} \sum_{l=0}^{L_p-1} h_{p,m}(l) u_m(n-l) + v(n), \quad 0 \leq n < N + L_p + L - 1, 1 \leq p \leq P, \quad (4)$$

where $M_T$ is the number of transmit antennas, $h_{p,m}(l) = g_p(l) h_m(l)$, $h_m(l)$ and $g_p(l)$ denote the complex RCS coefficient of the $l$th range cell for the $m$th waveform $u_m(n)$ and the $p$th spatial filtered response, respectively. The ideal form of the spatial filtered response [4] can be represented as

$$g_p(l) = \text{rect}\left(\frac{l - \sum_{i=0}^{p-1} L_i}{L_p}\right), \quad (5)$$

which is 1 in the interval $\left[\sum_{i=0}^{p-1} L_i, \sum_{i=0}^{p} L_i\right)$ and 0 elsewhere.

### III. MIMO SAR IRCI FREE RANGE RECONSTRUCTION

Following the range reconstruction algorithm for CP based OFDM SAR imaging for single transmitter [8], we remove the CP part in the discrete time received signal model in (4) by taking the $N$ samples starting from the $L$th sample point: for $0 \leq n < N$

$$z_p(n) = r_p(n + L)$$

$$= \sum_{m=1}^{M_T} \sum_{l=0}^{L_p-1} h_{p,m}(l) u_m(n + L - l) + v(n + L)$$

$$= \sum_{m=1}^{M_T} \sum_{l=0}^{L_p-1} h_{p,m}(l) s_m(n - l) + v(n + L). \quad (6)$$

Due to the CP insertion of $s_m(n)$, $s_m(n)$ in (6) is periodic with period $N$. Then, the $N$-point DFT of $z_p(n)$ becomes

$$Z_p(k) = \frac{1}{\sqrt{N}} \sum_{n=0}^{N-1} z_p(n) \exp\left(\frac{-j 2\pi nk}{N}\right) = \sqrt{N} \sum_{m=1}^{M_T} H_{p,m}(k) S_m(k) + V(k), \quad (7)$$

where $S_m(k)$ and $V(k)$ are the $N$-point DFTs of $s_m(n)$ and $v(n + L)$, $0 \leq n < N$, respectively, and

$$H_{p,m}(k) = \frac{1}{\sqrt{N}} \sum_{l=0}^{L_p-1} h_{p,m}(l) \exp\left(\frac{-j2\pi lk}{N}\right) = \frac{1}{\sqrt{N}} \sum_{n=0}^{N-1} \bar{h}_{p,m}(n) \exp\left(\frac{-j2\pi nk}{N}\right), \tag{8}$$

is the $N$-point DFTs of $\bar{h}_{p,m}(n)$ with

$$\bar{h}_{p,m}(n) = \begin{cases} h_{p,m}(n), & 0 \leq n < L_p \\ 0, & L_p \leq n < N \end{cases}. \tag{9}$$

In order to image the scatters in a swath, one needs to solve for the RCS coefficients $\bar{h}_{p,m}(n)$ that are involved in their DFT coefficients $H_{p,m}(k)$. Therefore, if one can solve for $H_{p,m}(k)$ from (7) for all indices $k$, the RCS coefficients $h_{p,m}(n)$ (equivalently $\bar{h}_{p,m}(n)$) can be solved and a swath can be imaged. From (7), one can see that for each $k$, there are $M_T$ variables $H_{p,m}(k)$ but there is only one equation, which looks like not enough. The idea in [9] is to produce $Q$ (at least $M_T$) equations as (7) by transmitting $Q$ consecutive OFDM pulses and then (7) would become a linear system of $Q$ equations with $M_T$ variables for each $k$. Although this is able to solve for these $M_T$ variables $H_{p,m}(k)$ for each $k$, the transmitter needs to transmit $Q$ consecutive OFDM pulses to obtain range profiles, which may be too long in time in applications to image a reasonable size swath.

The novelty of this paper is that we only use one OFDM pulse in imaging one swath as we have mentioned before, and solve for the RCS coefficients of a swath only based on one equation (7) for each $k$. The idea is that, since after the spatial filtering with multiple receive antennas, the RCS coefficients $\bar{h}_{p,m}(n)$ have the property (9), i.e., only $L_p$ of them are not zero, we will specially design the complex-valued weighting sequences $S_m(k)$ in (1) and (7) so that we can use both the frequency domain equations (7) and their corresponding time domain equations to solve for the RCS coefficients directly.

To do so, we first design that all the complex-valued weighting sequences $S_m(k)$ in (1) and (7) have constant module, say $|S_m(k)|=1$, for all $m$ and $k$, which is desired in the range reconstruction for CP based OFDM SAR imaging in order to have the SNR after the imaging maximized [8]. For convenience, in what follows, we only consider two transmit antennas, i.e., $M_T=2$.

We then apply the discrete frequency domain matched filter to (7) with the complex-valued weighting sequences $S_m(k)$ of the first waveform:

$$Y_{p,1}(k) = \frac{1}{\sqrt{N}} S_1^*(k) Z_p(k) = S_1^*(k) \sum_{m=1}^{2} H_{p,m}(k) S_m(k) + \frac{1}{\sqrt{N}} S_1^*(k) V(k)$$

$$= H_{p,1}(k) + H_{p,2}(k) G(k) + \bar{V}_1(k), \tag{10}$$

where * denotes complex conjugate, $G(k) = S_1^*(k)S_2(k)$ and $\bar{V}_1(k) = \frac{1}{\sqrt{N}}S_1^*(k)V(k)$. Then, taking the $N$-point IDFT of the two sides of (10) in terms of $k$, we obtain

$$\hat{h}_{p,1}(n) = \frac{1}{\sqrt{N}}\sum_{k=0}^{N-1}Y_{p,1}(k)\exp\left(\frac{j2\pi nk}{N}\right) = \bar{h}_{p,1}(n) + \bar{h}_{p,2}(n) \otimes g(n) + \bar{v}_1(n), \quad (11)$$

where $g(n)$ and $\bar{v}_1(n)$ are the $N$-point IDFTs of $G(k)$ and $\bar{V}_1(k)$, respectively, and $\otimes$ is the cyclic convolution. We notice that from (3) and (9), $\bar{h}_{p,m}(n)$ only have nonzero values at the first $L_p$ time indices $n$ and $L_p \leq N/2$. From (11), one can see that if $g(n)$ is designed such that the cyclic convolution $\bar{h}_{p,2}(n) \otimes g(n)$ becomes a cyclic shift of $\bar{h}_{p,2}(n)$ with the shift amount $N/2$, then, the two sets of non-zero RCS coefficients $\bar{h}_{p,1}(n)$ and $\bar{h}_{p,2}(n)$, will not overlap each other, i.e., IRCI free range profiles from any channel can be obtained. According to the property of the $N$-point DFT, we need to have $G(k) = S_1^*(k)S_2(k) = \exp(j\pi k)$, i.e.,

$$S_2(k) = S_1(k)\exp(j\pi k). \quad (12)$$

With this property of $g(n)$, from (11) we can get

$$\hat{h}_{p,1}(n) = \bar{h}_{p,1}(n) + \bar{h}_{p,2}\left(\left\langle n + \frac{N}{2}\right\rangle_N\right) + \bar{v}_1(n) = \frac{1}{\sqrt{N}}\sum_{k=0}^{N-1}Y_{p,1}(k)\exp\left(\frac{j2\pi nk}{N}\right), \quad (13)$$

where the notation $\langle x \rangle_y$ denotes the remainder of $x$ modulo $y$. Similarly, the other response can be obtained

$$\hat{h}_{p,2}(n) = \bar{h}_{p,2}(n) + \bar{h}_{p,1}\left(\left\langle n - \frac{N}{2}\right\rangle_N\right) + \bar{v}_2(n) = \frac{1}{\sqrt{N}}\sum_{k=0}^{N-1}Y_{p,2}(k)\exp\left(\frac{j2\pi nk}{N}\right), \quad (14)$$

where $\bar{v}_2(n)$ is the $N$-point IDFT of $\bar{V}_2(k)$. We then obtain $\bar{h}_{p,m}(n)$ from (13) and (14) as follows

$$\bar{h}_{p,m}(n) = \begin{cases} \hat{h}_{p,m}(n), & 0 \leq n < N/2 \\ 0, & N/2 \leq n < N \end{cases}. \quad (15)$$

From (9), the original subswath RCS coefficients $h_{p,m}(n)$ can be solved IRCI freely as:

$$h_{p,m}(n) = \bar{h}_{p,m}(n) = \frac{1}{\sqrt{N}}\sum_{k=0}^{N-1}Y_{p,m}(k)\exp\left(\frac{j2\pi nk}{N}\right), \quad 0 \leq n < L_p. \quad (16)$$

Then, the whole swath RCS coefficients $h_m(n)$ can be obtained

$$h_m(n) = [h_{1,m}(n), h_{2,m}(n), \cdots, h_{P,m}(n)], \quad m = 1, 2. \quad (17)$$

From the above range reconstruction, one can see that all the range cell scattering coefficients are recovered without any IRCI from other range cells, i.e., they are IRCI free. The proposed IRCI free range reconstruction algorithm diagram for each subswath is shown in Fig. 3.

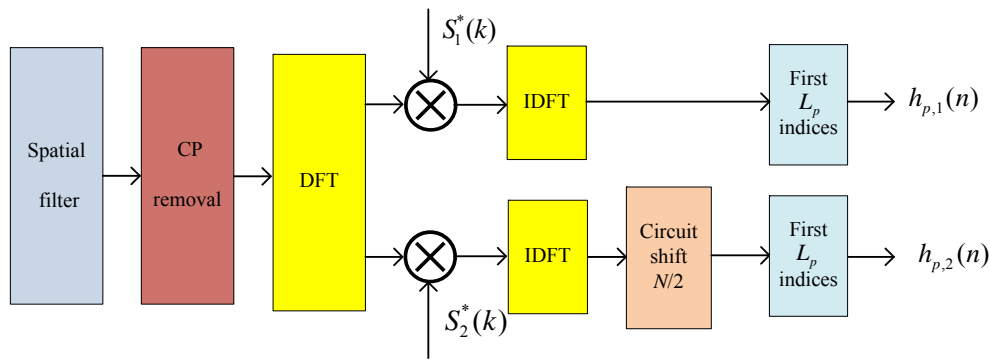

Fig. 3. MIMO radar IRCI free range reconstruction.

The proposed MIMO SAR scheme can be easily extended to more transmit antennas if (3) is met, which can be guaranteed by dividing the whole swath into more subswaths with increased receive antenna number. We can obtain the whole swath imaging by solving and putting together all the $P$ subswaths RCS coefficients $h_{p,m}(n)$ for all $n$, $0 \leq n < \sum_{p=1}^{P} L_p$ and all $m$, $1 \leq m \leq M_T$, which is IRCI free and has the full transmit spatial diversity.

We next design the two complex-valued weighting sequences $S_m(k)$ in (1) and (7) to satisfy the property (12) needed to shift the RCS coefficients from the two transmit antennas in one subswath non-overlapping from each other, and also the constant modular property used for the discrete frequency domain matched filtering to (7) to obtain (10). In addition, the orthogonality of the waveforms in discrete frequency domain and the low PAPR in discrete time domain need to be considered in the designs.

## IV. CIRCULARLY SHIFTED ZADOFF-CHU SEQUENCES

In this section, we propose to use circularly shifted Zadoff-Chu sequences as the complex-valued weighting sequences $S_m(k)$ that will satisfy (12) and have constant module.

In order to meet the aforementioned requirements, we analyze the property of circularly shifted Zadoff-Chu sequences. Let $S_1(k)$ be a Zadoff-Chu sequence,

$$S_1(k) = \exp\left(\frac{-j\pi\mu k\left(k + \langle N \rangle_2\right)}{N}\right), \quad (18)$$

where $\mu$ is an integer less than and relatively co-prime to $N$. The discrete time domain waveform of $S_1(k)$ by taking its $N$-point IDFT is (the similar derivation as [12])

$$s_1(n) = \frac{1}{\sqrt{N}} \sum_{k=0}^{N-1} S_1(k) \exp\left(\frac{j2\pi nk}{N}\right)$$

$$= S_1^*(\mu^{-1}n) \exp\left(\frac{-j2\pi n\langle N\rangle_2}{N}\right) s_1(0). \tag{19}$$

It can be seen from (19) that $|s_1(n)| = |s_1(0)|$ holds for every $n$, i.e., the IDFT of a Zadoff-Chu sequence is a constant modular sequence, i.e., the PAPR in the discrete time domain is optimal. For simplicity, let the sequence length $N$ be an even number and thus $\langle N\rangle_2 = 0$ in (18) and (19). In this case, (19) can be simplified as

$$s_1(n) = S_1^*(\mu^{-1}n) s_1(0). \tag{20}$$

The circularly shifted form of $S_1(k)$ with shift amount $N/2$ has the form

$$S_1\left(\left\langle k - \frac{N}{2}\right\rangle_N\right) = \beta S_1(k) \exp(j\pi\mu k), \tag{21}$$

where $\beta$ is a constant with $\beta = \exp(-j\pi\mu N/4)$. Since $\mu$ is an integer less than and relatively co-prime to $N$ and $N$ is even, $\mu$ must be an odd number. Thus,

$$\exp(j\pi\mu k) = \exp(j\pi k). \tag{22}$$

Let the complex-valued weighting sequence $S_2(k)$ take the following sequence

$$S_2(k) = \beta^* S_1\left(\left\langle k - \frac{N}{2}\right\rangle_N\right) = S_1(k) \exp(j\pi k). \tag{23}$$

Then, (12) is satisfied. From the zero periodic correlation property of a Zadoff-Chu sequence [11], we know that $S_1(k)$ and $S_2(k)$ are orthogonal. The advantage of this discrete frequency domain orthogonality is that it is not affected by any time delays in the time domain, while the discrete time domain orthogonality is sensitive to time delays.

According to the property of IDFT,

$$s_2(n) = \beta^* s_1(n) \exp(j\pi n), \tag{24}$$

which is also a constant modular sequence. By now, the orthogonality in discrete frequency domain, constant modular property in both discrete frequency and discrete time domains, and (12) are all satisfied for the MIMO-OFDM SAR radar.

## V. SIMULATION RESULTS

In this section, we present some simulations to illustrate the performance of our proposed method. We first show the performance of the proposed MIMO radar IRCI free range reconstruction with our designed circularly shifted Zadoff-Chu sequences. The simulations are carried out under the following assumptions [6]: use of the optimum spatial filter and the scattering point model. The optimal spatial filter means that the antenna pattern is ideally covering the angular range of interest and suppressing the sidelobe signal. There are 2 transmit antennas and the number of subcarriers is $N$=1024. Suppose that the maximum range cell number of all subswaths $L_o$ is 200 which is less than $N/2$.

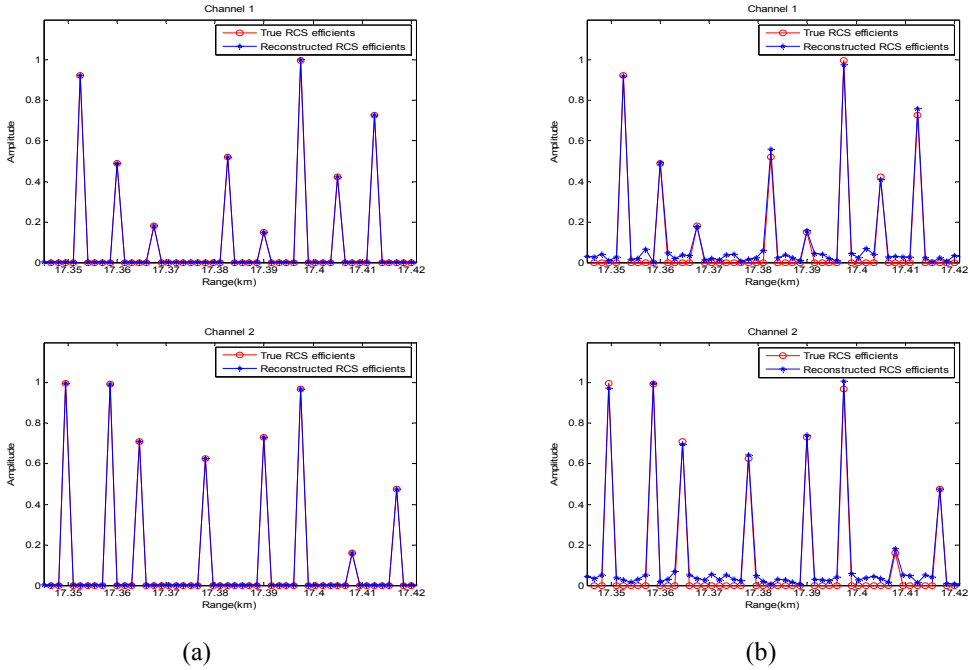

(a) (b)

Fig. 4. Two channel reconstructed RCS scattering coefficients: (a) without noise; (b) with noise.

Fig. 4(a) and Fig. 4(b) show the reconstructed RCS scattering coefficients from two transmit antennas without noise and with noise, respectively, using the circularly shifted Zadoff-Chu sequences and our IRCI free method. There are 8 strong scattering points for both channels and the scattering coefficients are randomly generated. Suppose that the signal-to-noise (SNR) is 0dB before the range reconstruction, where SNR refers to the ratio of the power of the strongest scattering point to that of noise at the receiver. Fig. 4(a) shows that both range profiles can be reconstructed perfectly without noise. It can be seen from Fig. 4(b) that the proposed range reconstruction method has a good performance in noisy environment without IRCI from any channel.

In order to demonstrate the low range sidelobes of the proposed method, we compare with the MIMO chirp waveform using the waveforms 3 and 4 in [5] and OFDM chirp waveform [6] with the matched filter. In Fig. 5, ○ denotes the amplitude of true range profile, +, * and x denote the reconstructed range profiles of the proposed method, MIMO chirp waveform and OFDM chirp waveform, respectively. Since for our proposed method there are no IRCI between scattering points in different range cells, the range profile can be recovered perfectly, which can be seen from Fig. 5(a). Fig. 5 (b) and Fig. 5 (c) are reconstructed range profiles of the MIMO chirp waveform and OFDM chirp waveform with matched filter, respectively. It can be seen that these two waveforms with matched filtering lead to some disadvantage, such as inaccurate peak amplitudes and invisible faint scatter points, which may damage the range profile. It is because that these two methods are not IRCI free, i.e., because of the result of the interaction of the sidelobes and mainlobes of closely-spaced scattering points.

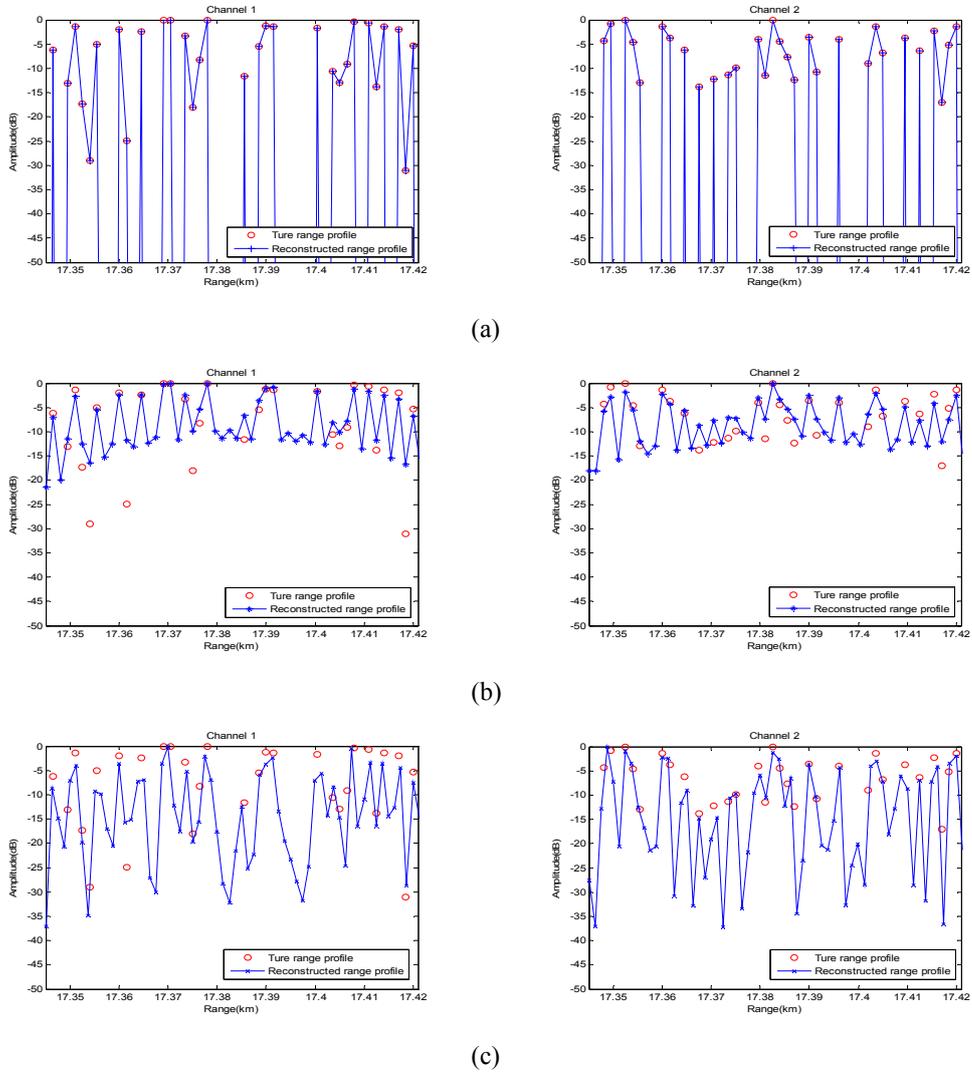

Fig. 5. Reconstructed range profiles: (a) the proposed method; (b) MIMO chirp waveform with matched filter; (c) OFDM chirp waveform with matched filter.

## VI. CONCLUSION

In this letter, we have proposed a CP based MIMO-OFDM SAR system, where each transmitter transmits a single OFDM pulse to obtain range profiles in a single swath and has the same signal bandwidth, i.e., the range resolution is not reduced. We have proposed to use circularly shifted Zadoff-Chu sequences as the weighting coefficients in the OFDM waveforms for different transmit antennas and apply spatial filters with multiple receive antennas to divide the whole swath into multiple subswaths, and then each subswath is reconstructed/imaged using our proposed IRCI free range reconstruction method. The multiple channel range profiles from different transmitters can be obtained separately without any IRCI, i.e., the spatial diversity can be achieved. By comparing with the MIMO chirp waveform and OFDM chirp waveform with the matched filter receiver, simulations are provided to illustrate the superior performance of the proposed IRIC free range reconstruction for MIMO-OFDM SAR.